\documentstyle[12pt,epsf,rotate,cite]{article}
\newcommand{\newsection}[1]{\section{#1}\setcounter{equation}{0}}

\newcommand{\be}{\begin{equation}}
\newcommand{\ee}{\end{equation}}
\newcommand{\bea}{\begin{eqnarray}}
\newcommand{\eea}{\end{eqnarray}}
\newcommand{\bd}{\begin{displaymath}}
\newcommand{\ed}{\end{displaymath}}
\newcommand{\f}{\frac}
\newcommand{\ra}{\rightarrow}
\newcommand{\me}[1]{\langle#1\rangle}
\newcommand{\al}{\alpha_s}
\newcommand{\aem}{\alpha}
\newcommand{\Bsee}{$B \ra X_s e^+ e^-$ }
\newcommand{\bsee}{$b \ra s e^+ e^-$ }
\newcommand{\bcenu}{$b \ra c e \bar\nu $ }
\newcommand{\kpiee}{$K_L \ra \pi^0 e^+ e^-$ }
\newcommand{\Lms}{\Lambda_{\overline{\rm MS}}}
\newcommand{\Ctilde}{\widetilde C}

\newcommand{\mt}{m_{\rm t}}
\newcommand{\mb}{m_{\rm b}}
\newcommand{\mc}{m_{\rm c}}
\newcommand{\ms}{m_{\rm s}}

\newcommand{\mw}{M_{\rm W}}
\newcommand{\mz}{M_{\rm Z}}

\newcommand{\gev}{\, {\rm GeV}}
\newcommand{\mev}{\, {\rm MeV}}

\newcommand{\RE}{{\rm Re}}

\begin{document}
\bibliographystyle{physics}
\renewcommand{\thefootnote}{\fnsymbol{footnote}}

\author{
Andrzej J. BURAS${}^{1,2}$ and Manfred M\"UNZ${}^{1}$\\
{\small\sl ${}^{1}$ Physik Department, Technische Universit\"at
M\"unchen, D-85748 Garching, Germany.}\\
{\small\sl ${}^{2}$ Max-Planck-Institut f\"ur Physik
                    -- Werner-Heisenberg-Institut,}\\
{\small\sl F\"ohringer Ring 6, D-80805 M\"unchen, Germany.}
}
\date{}

\title{
{\large\sf
\rightline{MPI-PhT/94-96}
\rightline{TUM-T31-82/94}
\rightline{hep-ph/9501281}
\rightline{December 1994}
}
\vspace{3cm}
\bigskip
\bigskip
{\LARGE\sf Effective Hamiltonian for \Bsee Beyond Leading Logarithms
in the NDR and HV Schemes}
\footnote{Supported by the German Bundesministerium f\"ur Forschung
und Technologie under contract 06 TM 743 and the CEC Science project
SC1-CT91-0729.}  }

\maketitle
\thispagestyle{empty}
\begin{abstract}
\noindent
We calculate the next-to-leading QCD corrections to the effective
Hamiltonian for \Bsee in the NDR and HV schemes. We give for the first
time analytic expressions for the Wilson Coefficient of the operator
$Q_9 = (\bar s b)_{V-A}(\bar e e)_V$ in the NDR and HV
schemes. Calculating the relevant matrix elements of local operators
in the spectator model we demonstrate the scheme independence of the
resulting short distance contribution to the physical amplitude.
Keeping consistently only leading and next-to-leading terms, we find
an analytic formula for the differential dilepton invariant mass
distribution in the spectator model. Numerical analysis of the $\mt$,
$\Lms$ and $\mu \approx {\cal O}(\mb)$ dependences of this formula is
presented. We compare our results with those given in the literature.
\end{abstract}

\newpage
\setcounter{page}{1}

\setcounter{footnote}{0}
\renewcommand{\thefootnote}{\arabic{footnote}}

\newsection{Introduction} \label{s.introd}

The rare decay \Bsee has been the subject of many theoretical studies
in the framework of the standard model and its extensions such as the
two Higgs doublet models and models involving supersymmetry
\cite{HWS:87, GSW:89, BBMR:91, AMM:91, JW:90, DPT:93, AGM:94, GIW:94}.
In particular the strong dependence of \Bsee on $\mt$ has been
stressed by Hou et.~al.~\cite{HWS:87}. It is clear that once \Bsee
has been observed, it will offer an useful test of the standard model
and of its extensions. To this end the relevant branching ratio, the
dilepton invariant mass distribution and other distributions of
interest should be calculated with sufficient precision. In particular
the QCD effects should be properly taken into account.

The central element in any analysis of \Bsee is the effective
Hamiltonian for $\Delta B=1$ decays relevant for scales $\mu \approx
{\cal O}(\mb)$ in which the short distance QCD effects are taken into
account in the framework of a renormalization group improved
perturbation theory. These short distance QCD effects have been
calculated over the last years with increasing precision by several
groups \cite{GSW:89,GDSN:89,CRV:91} culminating in a complete
next-to-leading QCD calculation presented by Misiak in
ref.~\cite{Mis:93} and very recently in a corrected version in
\cite{Mis:94}.

The actual calculation of \Bsee involves not only the evaluation of
Wilson coefficients of ten local operators (see (\ref{oper})) which
mix under renormalization but also the calculation of the
corresponding matrix elements of these operators relevant for
\mbox{\Bsee.} The latter part of the analysis can be done in the spectator
model, which, according to heavy quark effective theory, for B-decays
should offer a good approximation to QCD. One can also include the
non-perturbative ${\cal O}(1/\mb^2)$ corrections to the spectator
model which enhance the rate for \Bsee by roughly 10\%
\cite{FLS:94}. A realistic phenomenological analysis should also
include the long distance contributions which are mainly due to the
$J/\psi$ and $\psi'$ resonances \cite{LMS:89, DTP:89, DT:91}. Since in
this paper we are mainly interested in the next-to-leading short
distance QCD corrections to the spectator model we will not include
these complications in what follows.

It is well known that the Wilson coefficients of local operators
depend beyond the leading logarithmic approximation on the
renormalization scheme for operators, in particular on the treatment
of $\gamma_5$ in $D\not= 4$ dimensions. This dependence must be
cancelled by the scheme dependence present in the matrix elements of
operators so that the final decay amplitude does not depend on the
renormalization scheme. In the context of \Bsee this point has been
emphasized in particular by Grinstein et.~al.~\cite{GSW:89}. Other
examples such as $K \ra \pi\pi$, $K_{L,S} \ra \pi^0 e^+ e^-$, $B \ra
X_s \gamma$ can be found in refs.~\cite{BJL:93,BMMP:94,BLMM:94}. The
interesting feature of \Bsee as compared to decays such as $K \ra \pi
\pi$, is the fact that due to the ability of calculating reliably the
matrix elements of all operators contributing to this decay, the
cancellation of scheme dependence can be demonstrated in the actual
calculation of the short distance part of the physical amplitude.

Now all the existing calculations of \Bsee use the NDR renormalization
scheme (anticommuting $\gamma_5$ in $D \not= 4$ dimensions). Even if
arguments have been given, in particular in \cite{GSW:89} and
\cite{Mis:93}, how the cancellation of the scheme dependence in \Bsee
would take place, it is of interest to see this explicitly by
calculating this decay in two different renormalization schemes. In
addition, in view of the complexity of next-to-leading order (NLO)
calculations and the fact that the only complete NLO analysis of \Bsee
has been done by a single person, it is important to check the results
of refs.~\cite{Mis:93, Mis:94}.

Here we will present the calculations of the Wilson coefficients and matrix
elements relevant for \Bsee in two renormalization schemes (NDR and HV
\cite{HV:72}) demonstrating the scheme independence of the resulting
amplitude. Beside this the main results of our paper are as follows:
\begin{itemize}
\item
We give for the first time analytic NLO expressions for the Wilson
coefficient of the operator $Q_9 = (\bar s b)_{V-A} (\bar e e)_V$ in
the NDR and HV schemes.
\item
Calculating the matrix elements of local operators in the spectator
model we fully agree with Misiak's result for the dilepton
invariant mass distribution very recently given in \cite{Mis:94}.
\item
We find, that in the HV scheme the scheme dependent term in the matrix
elements (the so called $\xi$-term) receives in addition to
current-current contributions also contributions from QCD penguin
operators which are necessary for the cancellation of the scheme
dependence in the final amplitude. This should be compared with the
discussion of the scheme dependence given in refs.~\cite{GSW:89} and
\cite{Mis:93} where the $\xi$-term received only contributions from
current-current operators.
\item
We stress that in a consistent NLO analysis of the decay \Bsee, one
should on one hand calculate the Wilson coefficient of the operator
$Q_9 = (\bar s b)_{V-A} (\bar e e)_V$ including leading and
next-to-leading logarithms, but on the other hand only leading
logarithms should be kept in the remaining Wilson coefficients. Only
then a scheme independent amplitude can be obtained. This special
treatment of $Q_9$ is related to the fact that strictly speaking in
the leading logarithmic approximation only this operator contributes
to \Bsee. The contributions of the usual current-current operators,
QCD penguin operators, magnetic penguin operators and of $Q_{10} =
(\bar s b)_{V-A}(\bar e e)_A$ enter only at the NLO level and to be
consistent only the leading contributions to the corresponding Wilson
coefficients should be included. In this respect we differ from the
original analysis of Misiak \cite{Mis:93} who in his numerical
evaluation of \Bsee also included partially known NLO corrections to
Wilson coefficients of operators $Q_i (i\not= 9)$. These additional
corrections are, however, scheme dependent and are really a part of
still higher order in the renormalization group improved perturbation
theory. The most recent analysis of Misiak
\cite{Mis:94} does not include these contributions and can be directly
compared with the present paper.
\item
Keeping consistently only the leading and next-to-leading contributions to
\Bsee we are able to give analytic expressions for {\em all\/} Wilson
coefficients which should be useful for phenomenological applications.
\end{itemize}

\noindent
Our paper is organized as follows:

\noindent
In sect.~\ref{masterform} we collect the master formulae for \Bsee in
the spectator model which include consistently leading and
next-to-leading logrithms. In sect.~\ref{sectechnical} we describe
some details of the NLO calculation of the Wilson coefficient
$C_9(\mu)$ and of the relevant one-loop matrix elements in NDR and HV
schemes.  In sect.~\ref{secnumerics} we present a numerical
analysis. We end our paper with a brief summary of the main results.

\newsection{Master Formulae} \label{masterform}

\subsection{Operators} \label{opsec}
Our basis of operators is given as follows:
\be \label{oper}
\begin{array}{rcl}
Q_1    & = & (\bar{s}_{\alpha}  c_{\beta })_{V-A}
           (\bar{c}_{\beta }  b_{\alpha})_{V-A}       \vspace{0.2cm} \\
Q_2    & = & (\bar{s} c)_{V-A}  (\bar{c} b)_{V-A}     \vspace{0.2cm} \\
Q_3    & = & (\bar{s} b)_{V-A}\sum_q(\bar{q}q)_{V-A}  \vspace{0.2cm} \\
Q_4    & = & (\bar{s}_{\alpha}  b_{\beta })_{V-A} \sum_q (\bar{q}_{\beta}
          q_{\alpha})_{V-A}    \vspace{0.2cm} \\
Q_5    & = & (\bar{s} b)_{V-A}\sum_q(\bar{q}q)_{V+A}  \vspace{0.2cm} \\
Q_6    & = & (\bar{s}_{\alpha}  b_{\beta })_{V-A}
   \sum_q  (\bar{q}_{\beta }  q_{\alpha})_{V+A} \vspace{0.2cm} \\
Q_7    & = & \f{e}{8\pi^2} m_b \bar{s}_\alpha \sigma^{\mu\nu}
          (1+\gamma_5) b_\alpha F_{\mu\nu}            \vspace{0.2cm} \\
Q_8    & = & \f{g}{8\pi^2} m_b \bar{s}_\alpha \sigma^{\mu\nu}
   (1+\gamma_5)T^a_{\alpha\beta} b_\beta G^a_{\mu\nu} \vspace{0.2cm} \\
Q_9    & = & (\bar{s} b)_{V-A}  (\bar{e}e)_V          \vspace{0.2cm} \\
Q_{10} & = & (\bar{s} b)_{V-A}  (\bar{e}e)_A
\end{array}
\ee
where $\alpha$ and $\beta$ denote colour indices. We omit the colour
indices for the colour-singlet currents. Labels $(V \pm A)$ refer to
$\gamma_{\mu} (1 \pm \gamma_5)$. $Q_{1,2}$ are the current-current
operators, $Q_{3-6}$ the QCD penguin operators, $Q_{7,8}$ ``magnetic
penguin'' operators and $Q_{9,10}$ semi-leptonic electroweak penguin
operators. Our normalizations are as in refs.~\cite{BMMP:94} and
\cite{BLMM:94}.

\subsection{Wilson Coefficients} \label{coeffsect}

The Wilson coefficients for the operators $Q_1$--$Q_7$ are given in
the leading logarithmic approximation by \cite{BMMP:94, CFMRS:93,
CFMRS:94, CFRS:94}
\bea \label{coeffs}
C_j^{(0)}(\mu)    & = & \sum_{i=1}^8 k_{ji} \eta^{a_i}
  \qquad (j=1,...6) \vspace{0.2cm} \\
\label{C7eff}
C_7^{(0)eff}(\mu) & = & \eta^\f{16}{23} C_7^{(0)}(\mw) + \f{8}{3}
   \left(\eta^\f{14}{23} - \eta^\f{16}{23}\right) C_8^{(0)}(\mw) +
   \sum_{i=1}^8 h_i \eta^{a_i},
\eea
with
\bea
\eta & = & \f{\al(\mw)}{\al(\mu)}, \\
C_7^{(0)}(\mw) & = & - \f{1}{2} A(x_t), \\
C_8^{(0)}(\mw) & = & - \f{1}{2} F(x_t),
\eea
where $x_t = \mt^2/\mw^2$ and $A(x)$ and $F(x)$ are defined in
(\ref{A}) and (\ref{F}). The numbers $a_i$, $k_{ji}$ and $h_i$
 are given by
\be
\begin{array}{rrrrrrrrrl}
a_i = (\!\! & \f{14}{23}, & \f{16}{23}, & \f{6}{23}, & - \f{12}{23}, &
0.4086, & -0.4230, & -0.8994, & 0.1456 & \!\!)  \vspace{0.1cm} \\

k_{1i} = (\!\! & 0, & 0, & \f{1}{2}, & - \f{1}{2}, &
0, & 0, & 0, & 0 & \!\!)  \vspace{0.1cm} \\

k_{2i} = (\!\! & 0, & 0, & \f{1}{2}, &  \f{1}{2}, &
0, & 0, & 0, & 0 & \!\!)  \vspace{0.1cm} \\

k_{3i} = (\!\! & 0, & 0, & - \f{1}{14}, &  \f{1}{6}, &
0.0510, & - 0.1403, & - 0.0113, & 0.0054 & \!\!)  \vspace{0.1cm} \\

k_{4i} = (\!\! & 0, & 0, & - \f{1}{14}, &  - \f{1}{6}, &
0.0984, & 0.1214, & 0.0156, & 0.0026 & \!\!)  \vspace{0.1cm} \\

k_{5i} = (\!\! & 0, & 0, & 0, &  0, &
- 0.0397, & 0.0117, & - 0.0025, & 0.0304 & \!\!)  \vspace{0.1cm} \\

k_{6i} = (\!\! & 0, & 0, & 0, &  0, &
0.0335, & 0.0239, & - 0.0462, & -0.0112 & \!\!)  \vspace{0.1cm} \\

h_i = (\!\! & 2.2996, & - 1.0880, & - \f{3}{7}, & -
\f{1}{14}, & -0.6494, & -0.0380, & -0.0186, & -0.0057 & \!\!).
\end{array}
\ee

The first correct calculation of the two-loop anomalous dimensions
relevant for (\ref{C7eff}) has been presented in \cite{CFMRS:93,
CFMRS:94} and confirmed subsequently in \cite{CCRV:94b, CCRV:94a, Mis:94}.

The coefficient $C_8^{(0)eff}(\mu)$ does not enter the formula for
\Bsee at this level of accuracy. An analytic formula is given in
ref.~\cite{BMMP:94}.

The coefficient of $Q_{10}$ is given by
\be \label{C10}
C_{10}(\mw) =  \f{\aem}{2\pi} \Ctilde_{10}(\mw), \hspace{1cm}
\Ctilde_{10}(\mw) = - \f{Y(x_t)}{\sin^2\Theta_W}
\ee
with $Y(x)$ given in (\ref{YZ}). Since $Q_{10}$ does not renormalize
under QCD, its coefficient does not depend on $\mu\approx {\cal
O}(\mb)$. The only renormalization scale dependence in (\ref{C10})
enters through the definition of the top quark mass. We will return to
this issue in sect.~\ref{secnumerics}.

Finally, including leading as well as next-to-leading logarithms, we
find
\bea \label{C9}
C_9^{NDR}(\mu) & = & \f{\aem}{2\pi} \Ctilde_9^{NDR}(\mu) \\
\label{C9tilde}
\Ctilde_9^{NDR}(\mu) & = &
P_0^{NDR} + \f{Y(x_t)}{\sin^2\Theta_W} -4 Z(x_t) +
P_E E(x_t)
\eea
with
\bea
\label{P0NDR}
P_0^{NDR} & = & \f{\pi}{\alpha_s(M_W)} (-0.1875+ \sum_{i=1}^8 p_i
\eta^{a_i+1}) \nonumber \\
          &   & + 1.2468 +  \sum_{i=1}^8 \eta^{a_i} \lbrack
r^{NDR}_i+s_i \eta \rbrack \\
\label{PE}
P_E & = & 0.1405 +\sum_{i=1}^8 q_i\eta^{a_i+1} \\
\label{YZ}
Y(x) & = & C(x) - B(x), \hspace{1cm} Z(x) = C(x) + \f{1}{4} D(x).
\eea
Here
\bea
\label{A}
A(x) & = & \f{x(8x^2+5x-7)}{12(x-1)^3} + \f{x^2(2-3x)}{2(x-1)^4} \ln
x, \\
\label{B}
B(x) & = & \f{x}{4(1-x)} + \f{x}{4(x-1)^2} \ln x,\\
\label{C}
C(x) & = & \f{x(x-6)}{8(x-1)} + \f{x(3x+2)}{8(x-1)^2} \ln x, \\
\label{D}
D(x) & = & \f{-19 x^3 + 25 x^2}{36 (x-1)^3}
+ \f{x^2 (5 x^2 - 2 x - 6)}{18 (x-1)^4} \ln x- \f{4}{9} \ln x,\\
\label{E}
E(x) & = & \f{x (18 -11
x - x^2)}{12 (1-x)^3} + \f{x^2 (15 - 16 x + 4 x^2)}{6 (1-x)^4} \ln
x-\f{2}{3} \ln x, \\
\label{F}
F(x) & = & \f{x(x^2-5x-2)}{4(x-1)^3} + \f{3x^2}{2(x-1)^4} \ln x.
\eea

The coefficients $p_i$, $r^{NDR}_i$, $s_i$, and $q_i$ are found to
be as follows:
\be
\begin{array}{rrrrrrrrrl}
p_i = (\!\! & 0, & 0, & -\f{80}{203}, &  \f{8}{33}, &
0.0433, &  0.1384, & 0.1648 & - 0.0073 & \!\!)  \vspace{0.1cm} \\

r^{NDR}_{i} = (\!\! & 0, & 0, & 0.8966, & - 0.1960, &
- 0.2011, & 0.1328, & - 0.0292, & - 0.1858 & \!\!)  \vspace{0.1cm} \\

s_i = (\!\! & 0, & 0, & - 0.2009, &  -0.3579, &
0.0490, & - 0.3616, & -0.3554, & 0.0072 & \!\!)  \vspace{0.1cm} \\

q_i = (\!\! & 0, & 0, & 0, &  0, &
0.0318, & 0.0918, & - 0.2700, & 0.0059 & \!\!).
\end{array}
\ee

$P_E$ is ${\cal O}(10^{-2})$ and consequently the last term in
(\ref{C9tilde}) can be neglected. We keep it however in our
numerical analysis.

In the HV scheme only the coefficients $r_i$ are changed.
They are given by
\be
\begin{array}{lrrrrrrrrl}
r^{HV}_{i} = (\!\! & 0, & 0, & -0.1193, & 0.1003, &
- 0.0473, & 0.2323, & - 0.0133, & - 0.1799 & \!\!).
\end{array}
\ee
Equivalently we can write
\be \label{P0HV}
P_0^{HV} = P_0^{NDR} + \xi^{HV} \f{4}{9} \left( 3 C_1^{(0)} +
C_2^{(0)} - C_3^{(0)} -3 C_4^{(0)} \right)
\ee
with
\be \label{xi}
\xi = \left\{
\begin{array}{rl}
0,  & \mbox{NDR} \\
-1, & \mbox{HV.}
\end{array}
\right.
\ee
We note that
\bea
\label{sums1}
& & \sum_{i=1}^8 p_i = 0.1875, \hspace{4.4cm} \sum_{i=1}^8 q_i = -
0.1405, \\
\label{sums2}
& & \sum_{i=1}^8 (r_i + s_i) = - 1.2468 + \f{4}{9} (1 +
\xi), \hspace{1cm} \sum_{i=1}^8 p_i (a_i + 1) = - \f{16}{69}.
\eea
In this way for $\eta=1$ we find $P_E=0$, $P_0^{NDR} = 4/9$ and
$P_0^{HV} = 0$ in accordance with the initial conditions in
(\ref{C9prime}). Moreover, the second relation in (\ref{sums2})
assures the correct large logarithm in $P_0^{NDR}$, i.~e.~ $8/9\, \ln
(\mw/\mu)$. The derivation of (\ref{C9})--(\ref{P0HV}) is given in
sect.~\ref{sectechnical}.

\subsection{The Differential Decay Rate} \label{secdecayrate}
Introducing
\be \label{invleptmass}
\hat s = \f{(p_{e^+} + p_{e^-})^2}{\mb^2},\hspace{1cm} z =
\f{\mc}{\mb}
\ee
and calculating the one-loop matrix elements of $Q_i$ using the
spectator model in the NDR scheme we find
\bea \label{rate}
R(\hat s)\equiv\f{\f{d}{d\hat s} \Gamma (\mbox{\bsee})}{\Gamma
(\mbox{\bcenu})} & = & \f{\aem^2}{4\pi^2}
\left|\f{V_{ts}}{V_{cb}}\right|^2 \f{(1-\hat s)^2}{f(z)\kappa(z)} \cdot
\biggl[(1+2\hat s)\left(|\Ctilde_9^{eff}|^2 + |\Ctilde_{10}|^2\right)
+ \nonumber \\
& & 4 \left( 1 + \f{2}{\hat s}\right) |C_7^{(0)eff}|^2 + 12
C_7^{(0)eff} \ \RE\,\Ctilde_9^{eff}  \biggr]
\eea
where
\bea \label{C9eff}
\Ctilde_9^{eff} & = & \Ctilde_9^{NDR} \tilde\eta(\hat s) + h(z, \hat
s)\left( 3 C_1^{(0)} + C_2^{(0)} + 3 C_3^{(0)} + C_4^{(0)} + 3
C_5^{(0)} + C_6^{(0)} \right) \nonumber \\
& & - \f{1}{2} h(1, \hat s) \left( 4 C_3^{(0)} + 4 C_4^{(0)} + 3
C_5^{(0)} + C_6^{(0)} \right) \nonumber \\
& & - \f{1}{2} h(0, \hat s) \left( C_3^{(0)} + 3 C_4^{(0)} \right) +
\f{2}{9} \left( 3 C_3^{(0)} + C_4^{(0)} + 3 C_5^{(0)} + C_6^{(0)}
\right) .
\eea
Here
\bea \label{phasespace}
h(z, \hat s) & = & -\f{8}{9}\ln\f{m_b}{\mu} - \f{8}{9}\ln z +
\f{8}{27} + \f{4}{9} x \\
& & - \f{2}{9} (2+x) |1-x|^{1/2} \left\{
\begin{array}{ll}
\left( \ln\left| \f{\sqrt{1-x} + 1}{\sqrt{1-x} - 1}\right| - i\pi \right), &
\mbox{for } x \equiv \f{4z^2}{\hat s} < 1 \nonumber \\
2 \arctan \f{1}{\sqrt{x-1}}, & \mbox{for } x \equiv \f
{4z^2}{\hat s} > 1,
\end{array}
\right. \\
h(0, \hat s) & = & \f{8}{27} -\f{8}{9} \ln\f{\mb}{\mu} - \f{4}{9} \ln
\hat s + \f{4}{9} i\pi. \\
f(z) & = & 1 - 8 z^2 + 8 z^6 - z^8 - 24 z^4 \ln z, \\
\kappa(z)  & = & 1 - \f{2 \al(\mu)}{3\pi}\left[(\pi^2 - \f{31}{4})(1-z)^2
+ \f{3}{2} \right] \\
\tilde\eta(\hat s) & = & 1 + \f{\al(\mu)}{\pi}\, \omega(\hat s)
\eea
with
\bea \label{omega}
\omega(\hat s) & = & - \f{2}{9} \pi^2 - \f{4}{3}\mbox{Li}_2(s) - \f{2}{3}
\ln s \ln(1-s) - \f{5+4s}{3(1+2s)}\ln(1-s) - \nonumber \\
& &  \f{2 s (1+s) (1-2s)}{3(1-s)^2 (1+2s)} \ln s + \f{5+9s-6s^2}{6
(1-s) (1+2s)}.
\eea
Here $f(z)$ and $\kappa(z)$ are the phase-space factor and the single
gluon QCD correction to the \bcenu decay \cite{CM:79, KM:89}
respectively.  $\tilde\eta$ on the other hand represents single gluon
corrections to the matrix element of $Q_9$ with $\ms = 0$
\cite{JK:89,Mis:94}. For consistency reasons this correction should only
multiply the leading logarithmic term in $P_0^{NDR}$.

In the HV scheme the one-loop matrix elements are different and one
finds an additional explicit contribution to (\ref{C9eff}) given by
\be \label{MEHV}
- \xi^{HV} \f{4}{9} \left( 3 C_1^{(0)} + C_2^{(0)} - C_3^{(0)} - 3
C_4^{(0)} \right).
\ee
However $\Ctilde_9^{NDR}$ has to be replaced by $\Ctilde_9^{HV}$ given in
(\ref{C9tilde}) and (\ref{P0HV}) and consequently $\Ctilde_9^{eff}$ is the
same in both schemes.

The first term in the function $h(z, \hat s)$ in (\ref{phasespace})
represents the leading $\mu$-dependence in the matrix elements. It is
cancelled by the $\mu$-dependence present in the leading logarithm in
$\Ctilde_9$. The $\mu$-dependence present in the coefficients of the other
operators can only be cancelled by going to still higher order in the
renormalization group improved perturbation theory. To this end the
matrix elements of four-quark operators should be evaluated at
two-loop level. Also certain unknown three-loop anomalous dimensions
should be included in the evaluation of $C_7^{eff}$ and $C_9$
\cite{BMMP:94, BLMM:94}. Certainly this is beyond the scope of the
present paper and we will only investigate the left-over
$\mu$-dependence in sect.~\ref{secnumerics}.

The fact that the coefficient $C_9$ should include next-to-leading
logarithms and the other coefficients should be calculated in the
leading logarithmic approximation is easy to understand. There is a
large logarithm in $C_9$ represented by $1/\al$ in $P_0$ in
(\ref{P0NDR}). Consequently the renormalization group improved
perturbation theory for $C_9$ has the structure $ {\cal O}(1/\al) +
{\cal O}(1) + {\cal O}(\al)+ \ldots$ whereas the corresponding series
for the remaining coefficients is $ {\cal O}(1) + {\cal O}(\al)+
\ldots$. Therefore in order to find the next-to-leading ${\cal O}(1)$
term, the full two-loop renormalization group analysis for the operators
in (\ref{oper}) has to be performed in order to find $C_9$, but the
coefficients of the remaining operators should be taken in the leading
logarithmic approximation. This is gratifying because the coefficient
of the magnetic operator $Q_7$ is known only in the leading
logarithmic approximation. $Q_7$ does not mix with $Q_9$ and has no
impact on the coefficients $C_1$--$C_6$. Consequently the necessary
two-loop renormalization group analysis of $C_9$ can be performed
independently of the presence of the magnetic operators, which was
also the case of the decay \kpiee presented in ref.~\cite{BLMM:94}.

Let us finally compare our main formulae
(\ref{rate})--(\ref{MEHV}) with the ones given in the literature:
\begin{enumerate}
\item
The general expression (\ref{rate}) with $\kappa(z)=1$ is due to
Grinstein et.~al.~\cite{GSW:89} who in their approximate leading order
renormalization group analysis kept only the operators $Q_1, Q_2, Q_7,
Q_9, Q_{10}$.
\item
Inserting $C_i^{(0)}$ and $\Ctilde_9^{NDR}$ in (\ref{coeffs})and
(\ref{C10}) into (\ref{C9eff}) we find an analytic expression for
$\Ctilde_9^{eff}$ which agrees with a recent independent
calculation of Misiak \cite{Mis:94}.
\item
The sign of $i\pi$ in (\ref{phasespace}) differs from the one given in
\cite{GSW:89} and \cite{Mis:93} but agrees with \cite{Mis:94} and also
with the work of Fleischer \cite{Fleisch}.
\item
The ``$\xi$-term'' given in (\ref{MEHV}) contains in the HV scheme
also contributions from the operators $Q_3$ and $Q_4$, which are
however negligible. The discussion of the ``$\xi$-term''  in
refs.~\cite{GSW:89} and \cite{Mis:93} does not apply then to the HV
scheme.
\end{enumerate}

\newsection{Technical Details} \label{sectechnical}

\subsection{Wilson Coefficients}
In order to calculate the coefficient $C_9$ including next-to-leading
order corrections we have to perform in principle a two-loop
renormalization group analysis for the full set of operators given in
(\ref{oper}). However, $Q_{10}$ is not renormalized and the dimension
five operators $Q_7$ and $Q_8$ have no impact on $C_9$.  Consequently
only a set of seven operators, $Q_{1-6}$ and $Q_9$, has to be
considered. This is precisely the case of the decay \kpiee considered
in \cite{BLMM:94} except for an appropriate change of quark flavours
and the fact that now $\mu\approx {\cal O}(\mb)$ instead of
$\mu\approx {\cal O}(1\gev)$ should be considered. Because our
detailed NLO analysis of \kpiee has already been published we will
only discuss very briefly an analogous calculation of \Bsee, referring
the interested reader to \cite{BLMM:94}. We should stress that Misiak
\cite{Mis:93, Mis:94} used different conventions for the evanescent
operators than used in \cite{BLMM:94} and here. The agreement on
$\Ctilde_9^{eff}$ is therefore particularly satisfying.

Integrating out simultaneously $W, Z$ and $t$ we construct first the
effective Hamiltonian for $\Delta B=1$ transitions relevant for \bsee
with the operators normalized at $\mu = \mw$. Dropping the operators
$Q_7$, $Q_8$ and $Q_{10}$ for the reasons stated above and using the
unitarity of the CKM matrix we find
\bea \label{Heff}
{\cal H}_{eff}(\Delta B=1) & = & - \f{G_F}{\sqrt{2}} V_{ts}^* V_{tb} \left[
\sum_{i=1}^6 C_i(\mw) Q_i + C'_9(\mw) Q'_9 \right] \nonumber \\
& & + \f{G_F}{\sqrt{2}} V_{us}^* V_{ub} \left[ C_1(\mw) (Q_1^{(u)} -
Q_1) + C_2(\mw) (Q_2^{(u)} - Q_2) \right] .
\eea
Here $Q_{1,2}^{(u)}$ are obtained from $Q_{1,2}$ through the
replacement $c \ra u$. In order to make all the elements of the
anomalous dimension matrix be of the same order in $\al$, we have
appropriately rescaled $C_9$ and $Q_9$:
\be \label{rescaling}
Q'_9 = \f{\aem}{\al(\mu)} Q_9, \hspace{1cm} C'_9(\mu) =
\f{\al(\mu)}{\aem} C_9(\mu)
\ee
Note that because of GIM cancellation there are no penguin
contributions in the term proportional to $V_{us}^* V_{ub}$. They
would appear only at scales $\mu < \mc$ as was the case in
\mbox{\kpiee.} Since $|V_{us}^*V_{ub} / V_{ts}^* V_{tb}| < 0.02$ we
will drop the second term in what follows.

The initial conditions at $\mu=\mw$ for the coefficients $C_1$--$C_6$
in NDR and HV schemes have been given in sect.~2.4 and in the appendix
A of ref.~\cite{BLMM:94} respectively. Here it suffices to give only
the initial condition for the coefficient $C'_9$ (denoted by $C'_{7V}$
in
\cite{BLMM:94}) which reads:
\be \label{C9prime}
C'_9(\mw) = \f{\al(\mw)}{2\pi} \left[ \f{Y(x_t)}{\sin^2\Theta_W} - 4
Z(x_t) + \f{4}{9}(1+\xi) \right],
\ee
where $\xi$ has been defined in (\ref{xi}). The $x_t$ dependence
originates in box diagrams and in the $\gamma$- and $Z$-penguin
diagrams \cite{IL:81}.

With
\be \label{coeffvec}
\vec C^T \equiv \left( C_1, \ldots , C_6, C'_9 \right)
\ee
one can calculate the coefficients $C_i(\mu)$ by using the evolution
operator $\hat U_5(\mu,\mw)$ relevant for an effective theory with $f=5$
flavours:
\be \label{evol}
\vec C(\mu) = \hat U_5(\mu, \mw) \vec C(\mw).
\ee
An explicit expression for $\hat U_5$ is given in sect.~2 of \cite{BLMM:94}
where also the relevant expressions for one- and two-loop anomalous
dimensions can be found. One only has to set $f=5$, $u=2$ and $d=3$ in
the formulae given in \cite{BLMM:94}.

Using (\ref{evol}) and rescaling back the operator $Q_9$ we find at
$\mu\approx {\cal O}(\mb)$
\be \label{Heff_at_mu}
{\cal H}_{eff}(\Delta B=1) = - \f{G_F}{\sqrt{2}} V_{ts}^* V_{tb}
\left[ \sum_{i=1}^6 C_i(\mu) Q_i + C_9(\mu) Q_9 \right]
\ee
with the coefficient $C_9(\mu)$ given in (\ref{C9tilde}) and
(\ref{P0HV}) for NDR and HV schemes respectively. The result for HV
can either be found directly using (\ref{evol}) or by using the
relation
\be \label{NDRtoHV}
\vec C^{HV}(\mu) = \left( \hat 1 - \f{\al(\mu)}{4\pi} \Delta\hat r^T
\right) \vec C^{NDR}(\mu)
\ee
with the matrix $\Delta \hat r$ given in appendix A of
ref.\cite{BLMM:94}.

\subsection{One-Loop Matrix Elements}

\begin{figure}[ht]
\centerline{
\rotate[r]{
\epsfysize=6cm
\epsffile{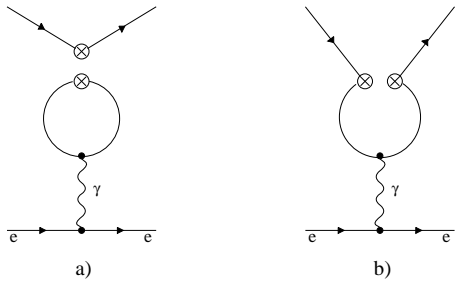}
}}
\vspace{0cm}
\caption[]{The two possibilites for insertion of a four-quark operator
into a penguin diagram.}
\label{fig1}
\end{figure}

The operators $Q_7$ and $Q_{10}$ contribute at this level of accuracy
only through tree level matrix elements. $Q_8$ contributes only
through the renormalization of $Q_7$ and its impact is only felt in
$C_7^{(0)eff}$. The four-quark operators $Q_{1-6}$, contribute at
one-loop level through the diagrams in fig.~\ref{fig1} where
``$\otimes\,\otimes$'' denotes the operator insertion. Finally at
next-to-leading level ${\cal O}(\al)$ corrections to the matrix
element $\me{Q_9}$ have to be calculated.

Let us begin with $\me{Q_{1-6}}$. As usual two types of insertions of
the operators into the penguin diagrams have to be considered. As
already discussed in ref.~\cite{BJLW:93} the appearance of a closed fermion
loop in fig.~\ref{fig1}a does not pose any problems in the NDR scheme
because nowhere in the calculation one has to evaluate Tr$[\gamma_\mu
\gamma_\nu \gamma_\rho \gamma_\lambda \gamma_5]$. The diagrams in
fig.~\ref{fig1} have been evaluated for the operators $Q_1$ and $Q_2$
by Grinstein et.~al.~\cite{GSW:89} and by Misiak \cite{Mis:93} for the
full set $Q_1$--$Q_6$. These calculations have been done in the NDR
scheme. Calculating these diagrams in the NDR and HV schemes we find
\bea \label{ME}
\me{Q_1} & = & \f{\aem}{2\pi} \left( 3 h(z,\hat s) - \f{4}{3}\xi
\right) \me{Q_9}_0 \nonumber \\
\me{Q_2} & = & \f{\aem}{2\pi} \left( h(z,\hat s) - \f{4}{9}\xi \right)
\me{Q_9}_0 \nonumber \\
\me{Q_3} & = & \f{\aem}{2\pi} \left( 3 h(z,\hat s) - 2 h(1, \hat s) -
\f{1}{2} h(0,\hat s) + \f{2}{3} + \f{4}{9}\xi \right) \me{Q_9}_0 \\
\me{Q_4} & = & \f{\aem}{2\pi} \left( h(z,\hat s) - 2 h(1, \hat s) -
\f{3}{2} h(0, \hat s) + \f{2}{9} +\f{4}{3}\xi \right) \me{Q_9}_0
\nonumber \\
\me{Q_5} & = & \f{\aem}{2\pi} \left( 3 h(z,\hat s) - \f{3}{2} h(1,
\hat s) + \f{2}{3} \right) \me{Q_9}_0 \nonumber \\
\me{Q_6} & = & \f{\aem}{2\pi} \left( h(z,\hat s) - \f{1}{2} h(1,
\hat s) + \f{2}{9} \right) \me{Q_9}_0 \nonumber
\eea
with $\xi$ defined in (\ref{xi}), $\me{Q_9}_0$ denoting the tree level
matrix element of $Q_9$ and
\be
h(z, \hat s) = \f{2}{3} G(z,\hat s) - \f{4}{9} - \f{8}{9} \ln \f{\mb}{\mu}.
\ee
Here
\be
G(z,\hat s) = -4 \int\limits_0^1 \!\!dx\, x(1-x) \ln \left( z^2 - \hat
s x(1-x) \right)
\ee
with $z$ and $\hat s$ defined in (\ref{invleptmass}).

A few remarks should be made:
\begin{itemize}
\item
$h(z, \hat s)$, $h(1, \hat s)$ and $h(0, \hat s)$ correspond to
internal $c$, $b$ and massless $(u,d,s)$ quarks in fig.~\ref{fig1}
respectively.
\item
The contributions of $(u,d,s)$ to diagram \ref{fig1}a) cancel each
other and consequently $h(0, \hat s)$ represents the contribution of
the internal strange quark in diagram \ref{fig1}b).
\item
We note that $\me{Q_5}$ and $\me{Q_6}$ matrix elements do not contain
the $\xi$-term. We should however stress that generally it is
certainly possible to find schemes in which $\me{Q_5}$ and $\me{Q_6}$
matrix elements can differ from the ones given in (\ref{ME}).
Similarly we have no argument that in schemes different from NDR and
HV the matrix elements are found simply by changing the value of $\xi$
in the formulae given above. It could be that the changes are more
involved. Consequently the discussions of the $\xi$-term presented in
\cite{GSW:89} and \cite{Mis:93} are not generally valid.
\end{itemize}

The one gluon correction to the matrix element of $Q_9$,
$\tilde\eta(\hat s)$, can be inferred from \cite{JK:89} as has been
noticed by Misiak in \cite{Mis:94}.
In \cite{JK:89} a left-handed current has been considered.  Thus we
rewrite the vector current as a sum of left- and right-handed
currents. Neglecting the electron masses these two contributions do
not interfere. Charge conjugation transforms the right-handed current
into a left-handed one. Since $\hat s$ is invariant under this
transformation both currents lead to the same invariant mass spectrum.
Therefore we can write
\be
\omega(\hat s) = \f{-2}{(1-\hat s)^2(1+2\hat s)} \int\limits_{\hat
s}^1\!\! dx \tilde F_1(x, \hat s)
\ee
with $\tilde F_1 (x, \hat s)$ defined explicitly in eq.~(3.9) of
\cite{JK:89}. Calculating the integral we arrive at the result given
in (\ref{omega}) which furthermore agrees with Misiak \cite{Mis:PC}.

\section{Numerical Analysis} \label{secnumerics}
In our numerical analysis we will use
\be
\al(\mu) = \f{4\pi}{\beta_0 \ln(\mu^2/\Lms^2)} \left[1 -
\f{\beta_1}{\beta_0^2} \f{\ln\ln(\mu^2/\Lms^2)}{\ln(\mu^2/\Lms^2)}
\right]
\ee
with $\beta_0 = 23/3$ and $\beta_1 = 116/3$ as appropriate for five
flavours. We also take $\Lms = (225 \pm 85) \mev$ corresponding to
$\al(\mz) = 0.117 \pm 0.007$. For the remaining parameters we take
\be
\begin{array}{rclrcl}
\aem & = & 1/129,\hspace{4cm} & \mc  & = & 1.4 \gev, \nonumber \\
\sin^2 \theta_W & = & 0.23, & \mb & = & 4.8 \gev, \nonumber \\
|V_{ts}/V_{cb}| & =&  1, & \mw  & = & 80.0 \gev.
\end{array}
\ee

In table \ref{tab1} we show the constant $P_0$ in (\ref{P0NDR}) for
different $\mu$ and $\Lms$, in the leading order corresponding to the
first term in (\ref{P0NDR}) and for the NDR and HV schemes as given by
(\ref{P0NDR}) and (\ref{P0HV}) respectively. In table \ref{tab2} we
show the corresponding values for $\Ctilde_9(\mu)$. To this end
we set $\mt= 170 \gev$.
\begin{table}[htb]
\begin{center}
\begin{tabular}{|c||c|c|c||c|c|c||c|c|c|}
\hline
& \multicolumn{3}{c||}{$\Lms = 0.140 \gev$} &
  \multicolumn{3}{c||}{$\Lms = 0.225 \gev$} &
  \multicolumn{3}{c| }{$\Lms = 0.310 \gev$} \\
\hline
$\mu [\gev]$ & LO & NDR & HV & LO & NDR & HV & LO & NDR & HV \\
\hline \hline
2.5 & 2.052 & 2.927 & 2.796 & 1.932 & 2.845 & 2.758 & 1.834 & 2.774 &
2.726 \\
5.0 & 1.851 & 2.623 & 2.402 & 1.787 & 2.589 & 2.394 & 1.735 & 2.560 &
2.387 \\
7.5 & 1.673 & 2.389 & 2.125 & 1.630 & 2.371 & 2.126 & 1.596 & 2.356 &
2.126 \\
10.0 & 1.524 & 2.202 & 1.910 & 1.493 & 2.192 & 1.915 & 1.468 & 2.183 &
1.919 \\
\hline
\end{tabular}
\end{center}
\vspace{-3mm}
\caption{The coefficient $P_0$ of $\widetilde C_9$ for various values
of $\Lms$ and $\mu$.}
\label{tab1}
\end{table}
\vspace{-4mm}
\begin{table}[htb]
\begin{center}
\begin{tabular}{|c||c|c|c||c|c|c||c|c|c|}
\hline
& \multicolumn{3}{c||}{$\Lms = 0.140 \gev$} &
  \multicolumn{3}{c||}{$\Lms = 0.225 \gev$} &
  \multicolumn{3}{c| }{$\Lms = 0.310 \gev$} \\
\hline
$\mu [\gev]$ & LO & NDR & HV & LO & NDR & HV & LO & NDR & HV \\
\hline \hline
2.5 & 2.052 & 4.495 & 4.364 & 1.932 & 4.413 & 4.326 & 1.834 & 4.341 &
4.293 \\
5.0 & 1.851 & 4.193 & 3.972 & 1.787 & 4.159 & 3.963 & 1.735 & 4.130 &
3.956 \\
7.5 & 1.673 & 3.960 & 3.696 & 1.630 & 3.942 & 3.696 & 1.596 & 3.926 &
3.697 \\
10.0 & 1.524 & 3.774 & 3.482 & 1.493 & 3.763 & 3.486 & 1.468 & 3.754 &
3.490 \\
\hline
\end{tabular}
\end{center}
\vspace{-3mm}
\caption{Wilson coefficient $\widetilde C_9$ for $\mt = 170 \gev$ and
various values of $\Lms$ and $\mu$.}
\label{tab2}
\end{table}

We observe:
\vspace{-1mm}
\begin{itemize}
\item
The NLO corrections to $P_0$ enhance this constant relatively to the
LO result by roughly 45\% and 35\% in the NDR and HV schemes
respectively. This enhancement is analogous to the one found in the
case of \kpiee.
\item
In calculating $P_0$ in the LO we have used $\al(\mu)$ at one-loop
level. Had we used the two-loop expression for $\al(\mu)$ we
would find for $\mu=5 \gev$ and $\Lms = 225 \mev$ the value $P_0^{LO}
\approx 1.98$. Consequently the NLO corrections would have smaller
impact. Ref.~\cite{GSW:89} including the next-to-leading term $4/9$
would find $P_0$ values roughly 20\% smaller than $P_0^{NDR}$ given in
tab.~\ref{tab1}.
\item
It is tempting to compare $P_0$ in table \ref{tab1} with that found
in the absence of QCD corrections. In the limit $\al \ra 0$ we find
$P_0^{NDR} = 8/9 \, \ln(\mw/\mu) + 4/9$ and $P_0^{HV} = 8/9\,
\ln(\mw/\mu)$ which for $\mu = 5 \gev$ give $P_0^{NDR} = 2.91$ and
$P_0^{HV} = 2.46$. Comparing these values with
table~\ref{tab1} we conclude that the QCD suppression of $P_0$ present in the
leading order approximation is considerably weakened in the NDR
treatment of $\gamma_5$ after the inclusion of NLO corrections. It is
essentially removed for $\mu > 5 \gev$ in the HV scheme.
\item
The NLO corrections to $\Ctilde_9$ which include also the $\mt$-dependent
contributions are large as seen in table \ref{tab2}.  The results in
HV and NDR schemes are by more than a factor of two larger than the
leading order result $\Ctilde_9 = P_0^{LO}$ which consistently should not
include $\mt$-contributions. This demonstrates very clearly the
necessity of NLO calculation which allow a consistent inclusion of the
important $\mt$-contributions. For the same set of parameters the
authors of ref.~\cite{GSW:89} would find $\Ctilde_9$ to be smaller than
$\Ctilde_9^{NDR}$ by 10--15\%.
\item
The $\mu$ and $\Lms$ dependences of $\Ctilde_9$ are quite weak. We
also find that the $\mt$ dependence of $\Ctilde_9$ is rather
weak. Varying $\mt$ between $150 \gev$ and $190 \gev$ changes
$\Ctilde_9$ by at most 10\%. This weak $\mt$ dependence of $\Ctilde_9$
originates in the partial cancellation of $\mt$ dependences between
$Y(x_t)$ and $Z(x_t)$ in (\ref{C9tilde}) as already seen in the case
of \kpiee. Finally, the difference between $\Ctilde_9^{NDR}$ and
$\Ctilde_9^{HV}$ is small and amounts to roughly 5\%.
\end{itemize}

\begin{figure}[hbt]
\centerline{
\epsfysize=9cm
\epsffile{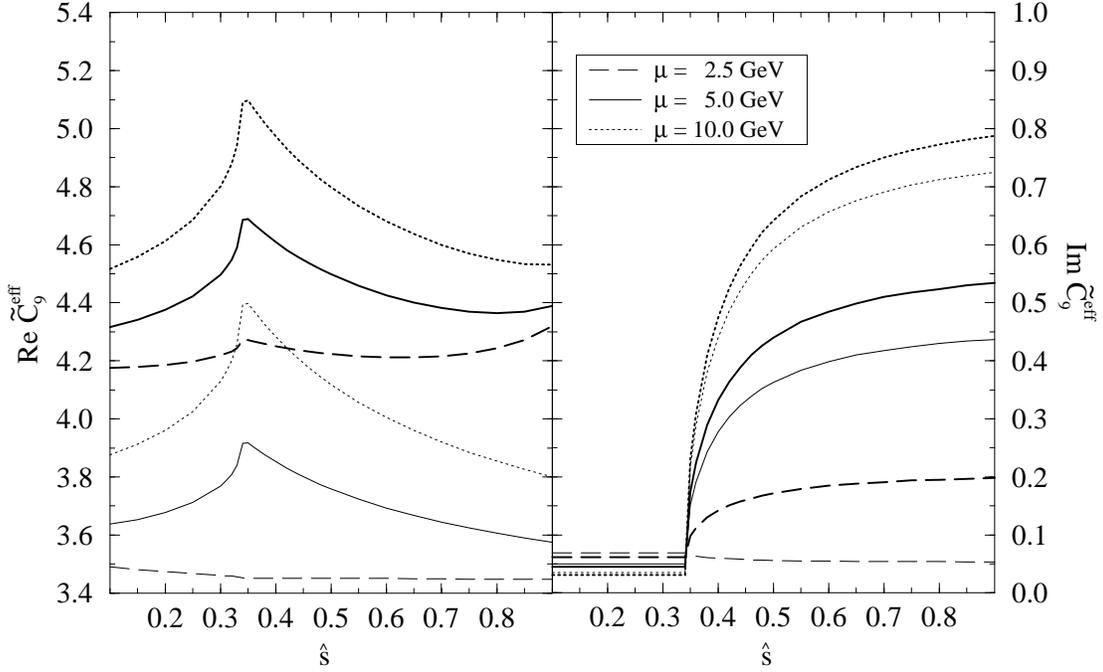}
}
\vspace{-2mm}
\caption[]{Comparison of the Wilson coefficient $\Ctilde_9^{eff}$ as a
function of $\hat s$ for $\mt = 170 \gev$, $\Lms = 0.225 \gev$ and
different values of $\mu$ in leading order (thin lines) and
next-to-leading order (thick lines) accuracy. Note the different
scales for the real and imaginary parts!}
\label{fig2}
\end{figure}
In fig.~\ref{fig2} we show $\Ctilde_9^{eff}$ of (\ref{C9eff}) as a function
of $\hat s$ for $\mt = 170 \gev$, $\Lms = 225 \mev$ and $2.5 \leq \mu
\leq 10 \gev$. In order to see the importance of the term resulting
from the one-loop matrix elements one should compare these results
with the $\hat s$-independent values of $\Ctilde_9$. We should also
remember that the NLO corrections to $P_0$ calculated here shift
$\Ctilde_9^{eff}$ for $\mu = 5.0 \gev$ by $\Delta \Ctilde_9^{NDR}
\approx 0.8$ and $\Delta \Ctilde_9^{HV} \approx 0.6$ with similar
results for other $\mu$. In order to show this effect more explicitly
we also plot in fig.~\ref{fig2} a ``leading order'' result obtained by
using only the leading term in (\ref{P0NDR}) with $\al$ at the
one-loop level but keeping otherwise all explicit NLO terms in
(\ref{C9tilde}) and the contributions from one-loop matrix elements
given in (\ref{C9eff}). It should be stressed that roughly 50\% of the
difference between the ``thick'' and ``thin'' lines in fig.~\ref{fig2}
is due to the term $4/9$ in (\ref{C9prime}) which in the NDR scheme
enters the NLO terms in $P_0$ but in the HV scheme is present in the
one-loop matrix elements. We have left it out in the ``thin'' lines in
fig.~\ref{fig2} in order to show its importance. The calculation of
NLO corrections to $P_0$ allows a consistent inclusion of this term
which contributes positively to $\Ctilde_9^{eff}$. Additional
enhancement comes from using the two-loop renormalization group
analysis for $\Ctilde_9$ and $\al$ at the two-loop level. In
fig.~\ref{fig2} we also note that $\mbox{Re} \Ctilde_9^{eff} \gg
\mbox{Im} \Ctilde_9^{eff}$. The pronounced peak for $\hat s = 4
\mc^2/\mb^2 = 0.34$ is related to the behaviour of $h(z,\hat s)$ in
(\ref{phasespace}). This peak essentially disappears for $\mu = 2.5
\gev$ because of the accidential cancellation $3 C_1^{(0)} + C_2^{(0)}
\approx 0$ in the dominant term multiplying $h(z, \hat s)$. The
authors of ref.~\cite{GSW:89} would find Re$\,\Ctilde_9^{eff}$ by
about 15\% below our values. In the absence of QCD corrections,
$h(z,\hat s)$ in (\ref{C9eff}) is multiplied by $C_2^{(0)} = 1$ and
consequently there is no accidental suppression of this term as in the
QCD case. Since in addition for $\al \ra 0$ \ $P_0^{NDR}$ is slightly
enhanced over the values given in table \ref{tab1}, we find
$\Ctilde_9^{eff}$ in the absence of QCD corrections to be
substantially larger than the result given in fig.~\ref{fig2}. For
instance, Re\,$\Ctilde_9^{eff}$ varies between 5.2 and 6.3 for $0.1
\leq \hat s \leq 0.9$. The complete result for $R(\hat s)$ in this
case is shown in fig.~\ref{fig6} at the end of this section.

\begin{figure}[htb]
\centerline{
\epsfysize=9cm
\epsffile{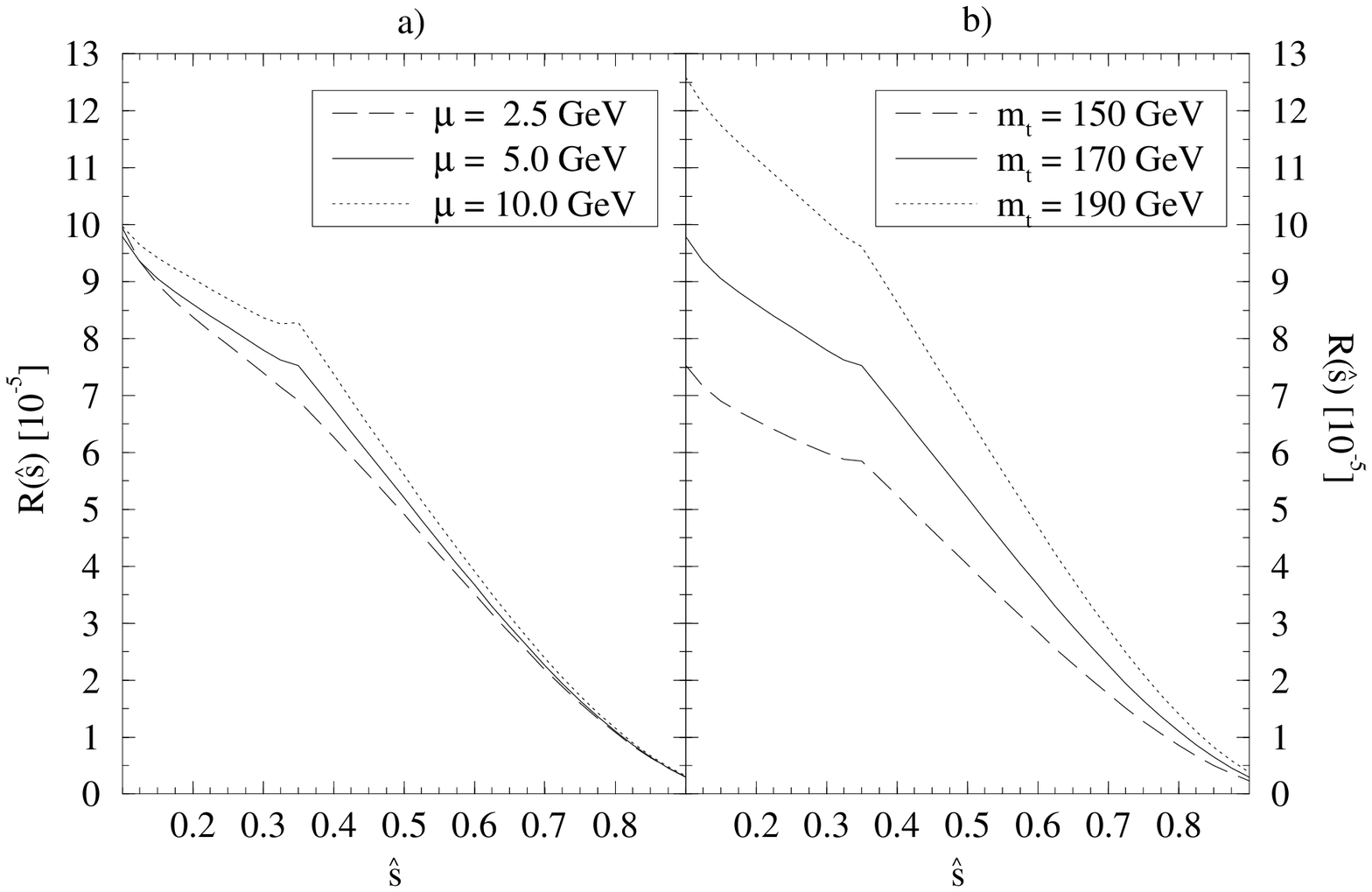}
}
\vspace{-2mm}
\caption[]{a) $R(\hat s)$ for $\mt = 170 \gev$, $\Lms = 225 \mev$ and
different values of $\mu$.\\ \phantom{Figure3: } b) $R(\hat s)$ for
$\mu = 5 \gev$, $\Lms = 225 \mev$ and various values of $\mt$.}
\label{fig3}
\end{figure}

We next present a numerical analysis of (\ref{rate}). In doing this we
keep in mind that for $\hat s \approx m_\psi^2 / \mb^2$, $\hat s
\approx m_{\psi'}^2 / \mb^2$ etc.~the spectator model cannot be the
full story and additional long distance contributions discussed in
refs.~\cite{LMS:89, DTP:89, DT:91} have to be taken into account in a
phenomenological analysis. Similarly we do not include $1/\mb^2$
corrections calculated in \cite{FLS:94} which typically enhance the
differential rate by about 10\%.


In fig.~\ref{fig3}a) we show $R(\hat s)$ for $\mt = 170 \gev$, $\Lms =
225 \mev$ and different values of $\mu$. In fig.~\ref{fig3}b) we set
$\mu = 5 \gev$ and vary $\mt$ from $150 \gev$ to $190 \gev$. The
remaining $\mu$ dependence is rather weak and amounts to at most $\pm
8\%$ in the full range of parameters considered. The $\mt$ dependence
of $R(\hat s)$ is sizeable. Varying $\mt$ between 150 GeV and 190 GeV
changes $R(\hat s)$ by typically 60--65\% which in this range of $\mt$
corresponds to $R(\hat s) \sim \mt^2$. It is easy to verify that this
strong $\mt$ dependence originates in the coefficient $\Ctilde_{10}$
given in (\ref{C10}) as already stressed by several authors in the past
\cite{HWS:87, GSW:89, BBMR:91, DPT:93, GIW:94, AGM:94, AMM:91, JW:90}.

We do not show the $\Lms$ dependence as it is very weak. Typically,
changing $\Lms$ from $140\mev$ to $310\mev$ decreases $R(\hat s)$ by
about 5\%.

\begin{figure}[htb]
\centerline{
\epsfysize=10cm
\epsffile{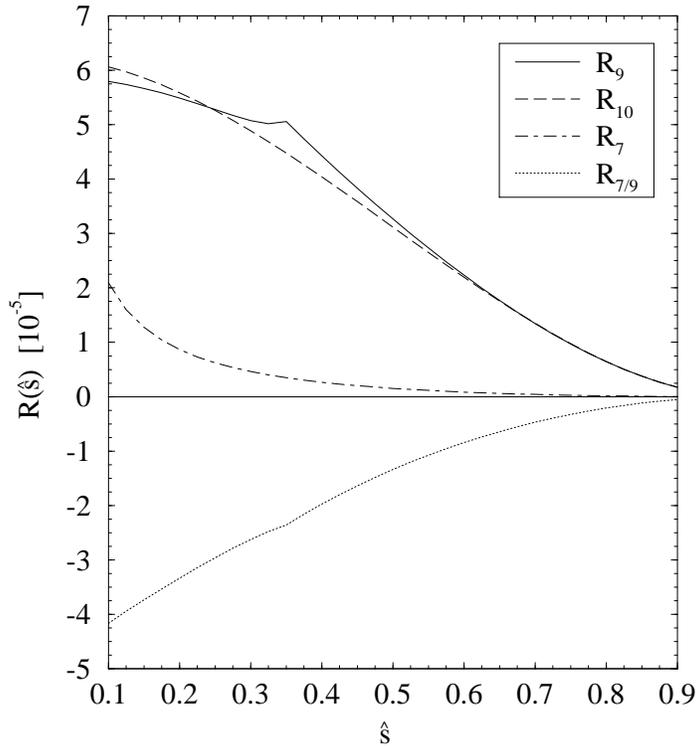}
}
\vspace{-2mm}
\caption[]{Comparison of the four different contributions to $R(\hat
s)$ according to eqn.~(\ref{rate}).}
\label{fig5}
\end{figure}

\begin{samepage}
$R(\hat s)$ is governed by three coefficients, $\Ctilde_9^{eff}$,
$\Ctilde_{10}$ and $C_7^{(0)eff}$. It is of interest to investigate
the importance of various contributions. To this end we set $\Lms =
225 \gev$, $\mt = 170 \gev$ and $\mu = 5 \gev$. In fig.~\ref{fig5} we
show $R(\hat s)$ keeping only $\Ctilde_9^{eff}$, $\Ctilde_{10}$,
$C_7^{(0)eff}$ and the $C_7^{(0)eff}$--$\Ctilde_9^{eff}$ interference
term, respectively. Denoting these contributions by $R_9$, $R_{10}$,
$R_7$ and $R_{7/9}$ we observe that the term $R_7$ plays only a minor
role in $R(\hat s)$. On the other hand the presence of $C_7^{(0)eff}$
cannot be ignored because the interference term $R_{7/9}$ is
significant. In fact the presence of this large interference term
could be used to measure experimentally the relative sign of
$C_7^{(0)eff}$ and $\mbox{Re}\,\Ctilde_9^{eff}$ \cite{GSW:89, AMM:91,
JW:90, GIW:94, AGM:94} which as seen in fig.~\ref{fig5} is negative in
the Standard Model. However, the most important contributions are
$R_9$ and $R_{10}$ in the full range of $\hat s$ considered. For $\mt
\approx 170 \gev$ these two contributions are roughly of the same
size. Due to a strong $\mt$ dependence of $R_{10}$, this
contribution dominates for higher values of $\mt$ and is less
important than $R_9$ for $\mt < 170 \gev$.
\end{samepage}

\begin{figure}[htb]
\centerline{
\epsfysize=10cm
\epsffile{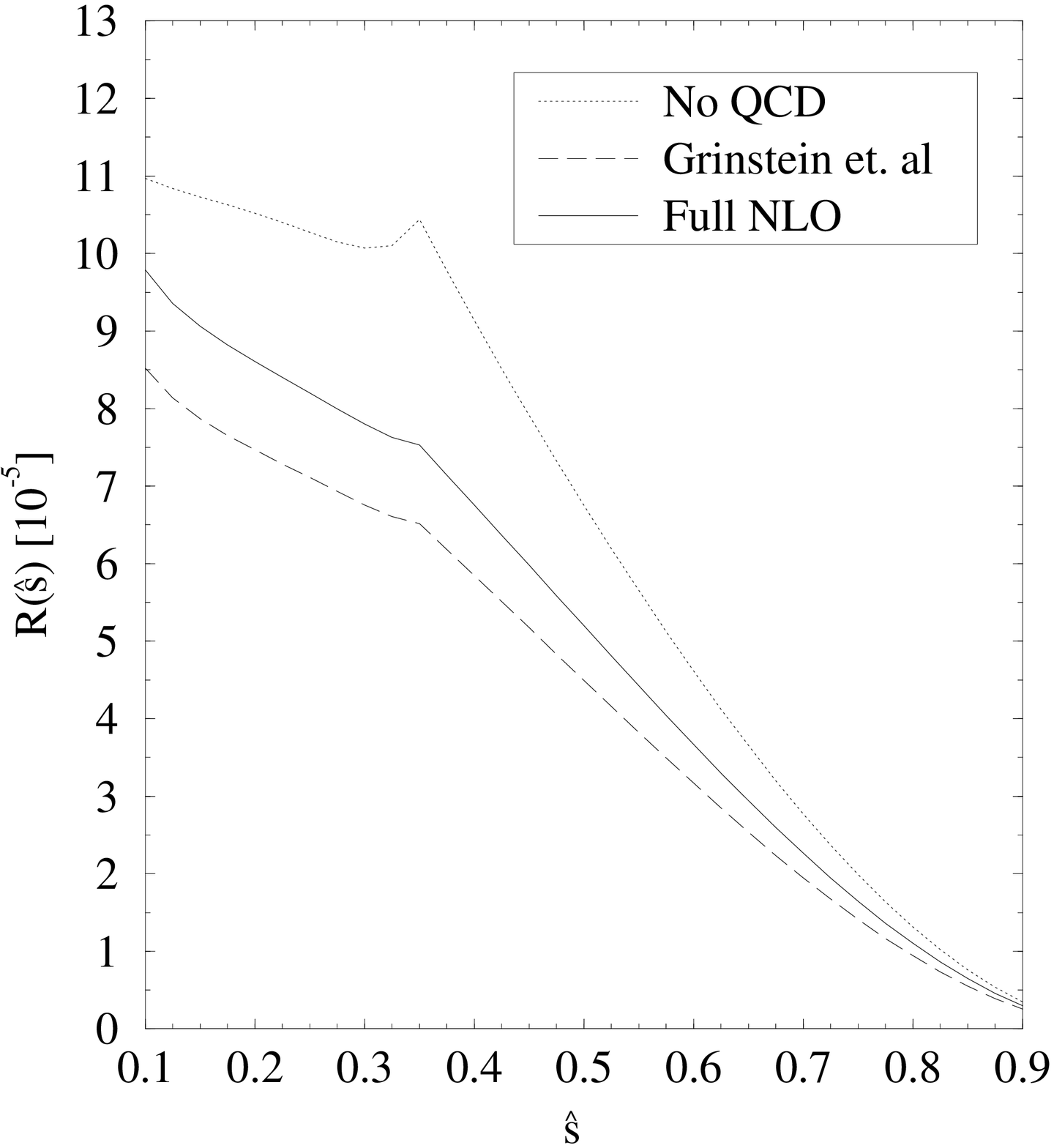}
}
\vspace{-2mm}
\caption[]{$R(\hat s)$ for $\mt = 170 \gev$, $\Lms = 225 \mev$ and
$\mu = 5 \gev$.}
\label{fig6}
\end{figure}

Next, in fig.~\ref{fig6} we show $R(\hat s)$ for $\mu = 5 \gev$,
$\mt = 170 \gev$ and $\Lms = 225 \mev$ compared to the case of no QCD
corrections and to the results Grinstein et.~al.~\cite{GSW:89} would
obtain for our set of parameters using their approximate leading order
formulae.

Finally, we would like to address the question of the definition of
$\mt$ used here. In order to be able to analyze this question, one
would have to calculate perturbative QCD corrections to the functions
$Y(x_t)$ and $Z(x_t)$ and include also an additional order in the
renormalization group improved perturbative calculation of $P_0$. The
latter would require evaluation of three-loop anomalous dimension
matrices, which in the near future nobody will attempt. In any case,
we expect only a small correction to $P_0$. The uncertainty due to the
choice of $\mu$ in $\mt(\mu)$ can be substantial, as stressed in
refs.~\cite{BB:93a,BB:93b}, and may result in 20--30\% uncertainties
in the branching ratios. It can only be reduced if ${\cal O}(\al)$
corrections to $Y(x_t)$ and $Z(x_t)$ are included. For
$K^+\ra\pi^+\nu\bar{\nu}$, $K_L\ra\pi^0\nu\bar{\nu}$, $B\ra\mu^+\mu^-$
and $B\ra X_s\nu\bar{\nu}$ this has been done in refs.~\cite{BB:93a,
BB:93b}. The inclusion of these corrections reduces the uncertainty in
the corresponding branching ratios to a few percent. Fortunately, the
result for the corrected function $Y(x_t)$ given in
refs.~\cite{BB:93a, BB:93b} can be directly used here. The message of
refs.~\cite{BB:93a, BB:93b} is the following: For $\mt =
\overline{m}_{\rm t}(\mt)$, the QCD corrections to $Y(x_t)$ and
consequently to $\Ctilde_{10}$ are below 2\%. Corresponding
corrections to $Z(x_t)$ are not known. Fortunately, the $\mt$
dependence of $\Ctilde_9$ is much weaker and the uncertainty due to
the choice of $\mu$ in $\mt(\mu)$ is small. On the basis of these
arguments and the result of refs.~\cite{BB:93a,BB:93b} we believe that
if $\mt = \overline{m}_{\rm t}(\mt)$ is chosen, the additional short
distance QCD corrections to BR(\Bsee) should be small.

\section{Summary}
We have calculated the effective Hamiltonian relevant for the rare
decay \Bsee beyond the leading logarithmic approximation. The main new
result of this paper is the calculation of the Wilson coefficient of
the operator $Q_9 = (\bar s b)_{V-A} (\bar e e)_V$ including
next-to-leading logarithms in the NDR and HV renormalization
schemes. A separate analytic expression for $C_9$ given in
sect.~\ref{masterform} as opposed to $C_9^{eff}$ given in
\cite{Mis:94} should be useful not only in \Bsee but also in $B \ra
K^* e^+ e^-$ and other rare $B$-decays to which $Q_9$ contributes.
Calculating \Bsee in the spectator model we confirm the very recent
result for $C_9^{eff}$ presented by Misiak in \cite{Mis:94}. The
cancellation of the scheme dependence in $C_9^{eff}$ is shown
explicitly in our paper.

The effect of the NLO corrections is to enhance BR(\Bsee$\!$) so that its
suppression found in the leading order analysis of ref.~\cite{GSW:89}
is considerably weakened. This is seen in particular in
fig.~\ref{fig6}.

We have investigated the $\mt$, $\Lms$ and $\mu \approx {\cal O}(\mb)$
dependence of the ``reduced'' branching ratio $R(\hat s)$. The
dependences on $\Lms$ and $\mu$ are rather small, at most $\pm 8\%$ in
the full range of parameters considered. The dependence on $\mt$ is
sizeable. In the range $150 \gev \leq \mt \leq 190 \gev$ it is roughly
parametrized by $R(\hat s) \sim \mt^2$. For $\mt = 170 \gev$, $\Lms =
225 \mev$, $\mu = 5 \gev$ and $0.1 \leq \hat s \leq 0.8$ we find
\be
1.0 \cdot 10^{-5} \leq R(\hat s) \leq 9.8 \cdot 10^{-5}.
\ee
This result can be modified by non-perturbative $1/\mb^2$ corrections
and long distance contributions \cite{LMS:89, DTP:89, DT:91}, which
are however beyond the scope of this paper.

\bigskip
\noindent
{\large\bf Acknowledgment}

\noindent
We would like to thank Miko{\l}aj Misiak for the correspondence
related to his independent analysis presented in ref.~\cite{Mis:94}.
One of us (M.M.) appreciates helpful discussions with M. Jamin, M.
Lautenbacher and U. Nierste.
\bigskip


\end{document}